# Refractive Index Robustness of Metalenses

DONGYOUNG LEE and JISOO KYOUNG*

*Department of Physics, Dankook University, 119 Dandae-ro, Dongnam-gu, Cheonan-si, Chungnam, 31116, Republic of Korea*
*\* [kyoungjs@dankook.ac.kr](mailto:kyoungjs@dankook.ac.kr)*

**Abstract:** Metalenses have emerged as a powerful platform for compact wavefront engineering; however, their performance stability under refractive index fluctuations induced by environmental perturbations, such as temperature shifts, remains a critical concern. Here, we demonstrate the intrinsic refractive index robustness of dielectric metalenses and elucidate its physical origin. By parametrically sweeping the refractive index, we observe that the metalens maintains a stable focal profile with negligible deviations in best-focus position and spot size over a broad range of variations. We identify that this robustness arises from the structural invariance of the zone boundaries: despite index-induced local phase deformations, the spatial locations of the $2\pi$ phase-reset boundaries remain stationary, thereby maintaining the effective wavefront gradient. Furthermore, we reveal that this robustness enables a "quasi-scale-invariant" focusing behavior, where the focusing performance follows a predictable linear trend under uniform geometric scaling even in the presence of material dispersion. Our findings suggest that metalenses can maintain stable focusing behavior against refractive index variations that may arise from unavoidable environmental perturbations in practical optical systems.

## 1. Introduction

A metalens is a planar metasurface-based lens composed of a two-dimensional array of subwavelength nanostructures, commonly referred to as meta-atoms. While conventional refractive lenses achieve focusing through the gradual accumulation of optical phase along the propagation direction via their thickness, metalenses directly reconstruct the wavefront at the surface, enabling lens functionality within an ultrathin form factor. Consequently, metalenses offer key advantages such as compactness, lightweight design, and compatibility with on-chip integration [1–4], while also allowing large-area fabrication and integration through lithography-based wafer-scale processes [5–9]. Furthermore, owing to their high design flexibility, metalenses can achieve extraordinarily high numerical apertures (NA) and broadband achromatic focusing within a single platform [10–13]. In practice, such high-efficiency and broadband performance is predominantly realized using dielectric metasurfaces, as plasmonic implementations suffer from intrinsic ohmic losses that limit transmission efficiency [14, 15]. Such versatility has driven the rapid expansion of metalens research into a wide range of applications, including imaging optics, AR/VR (Augmented Reality, Virtual Reality) systems, compact spectroscopy, integrated photonics, and optical sensing [4,16–19]. The fundamental operating principle of a metalens is to implement, on a planar surface, a phase profile corresponding to a target wavefront, where the local phase at each position is determined by the geometry or material properties of the constituent meta-atoms. Among these parameters, the refractive index is one of the key physical quantities governing the phase shift of each meta-atom: it modifies both the optical path length and resonance conditions inside the structure, while simultaneously affecting the transmission response into the surrounding medium [20]. As a result, variations in refractive index lead to changes in both transmission and phase shift even for identical geometries [21], which can cause deviations from the intended wavefront and ultimately induce functional changes or performance degradation.

The refractive index is not a fixed material parameter under practical operating conditions but can be modulated by various external factors. Temperature variations alter the refractive index through the thermo-optic effect [22,23], while mechanical stress and strain modify both the refractive index and birefringence via the photoelastic effect [24,25]. Additionally, under high-intensity optical fields, the refractive index may exhibit nonlinear modifications due to the optical Kerr effect [26,27]. Such refractive index fluctuations simultaneously perturb the phase shift and transmission of individual meta-atoms, potentially accumulating deviations from the designed wavefront and naturally leading to performance degradation in metalenses. Interestingly, however, in our previous work [28], the performance variations induced by temperature changes were found not to be dominated by refractive index variations; instead, substrate thermal expansion was found to alter the phase distribution through geometry changes, serving as the primary mechanism. This observation suggests that refractive index variations in metalenses may not directly translate into performance degradation to the extent intuitively expected. Consequently, a natural question arises: to what extent can metalens performance be preserved under refractive index variations, and what physical mechanisms underlie such robustness? A useful clue to this question comes from Arbabi et al. [29], who showed that chromatic dispersion in metalenses mainly originates from phase discontinuities at Fresnel-zone boundaries (where the phase undergoes a $2\pi$ reset). This occurs because the actual phase profile at an off-design wavelength remains closer to the design-wavelength phase distribution than to the ideal profile required at the new wavelength. Paradoxically, this suggests that the same phase-reset structure may, under fixed-wavelength refractive-index perturbations, help preserve the dominant focusing behavior examined in this work.

Since the refractive index of meta-atom materials can vary with operating conditions such as temperature, understanding how such variations influence metalens performance is important for the practical design and reliable operation of metalens-based optical devices. In this work, we parametrically vary the refractive index of the meta-atoms and quantitatively evaluate the resulting changes in focusing performance, with primary emphasis on best-focus position shift and spot size preservation, thereby elucidating the refractive index robustness of metalenses. Furthermore, we interpret why a metalens can relatively maintain its performance under refractive index variations by linking the deformation of the designed phase profile to the corresponding changes in focusing behavior, and we present the physical mechanism responsible for the observed robustness.

## 2. Methods

Numerical simulations were conducted using the finite difference time domain (FDTD) method (Ansys Lumerical). The unit cell characteristics were analyzed using periodic boundary conditions, while the full metalens simulations employed perfectly matched layer (PML) boundaries. The refractive indices of both $Si_3N_4$ and $SiO_2$ were modeled using the Sellmeier equation, and a linearly polarized plane wave at $\lambda_0 = 633$ nm was used as the source. For computational efficiency, the electric field distributions in the focal region were derived by projecting the near-field data recorded at $z = 1.02H$ (30 nm above the metalens) using a vectorial near-to-far field transformation. The best-focus position was determined by the position of maximum intensity along the optical axis, and the spot size was defined as the full width at half maximum (FWHM) at the best-focus plane.

## 3. Results and Discussion

3.1 Design and refractive-index-dependent focusing behavior

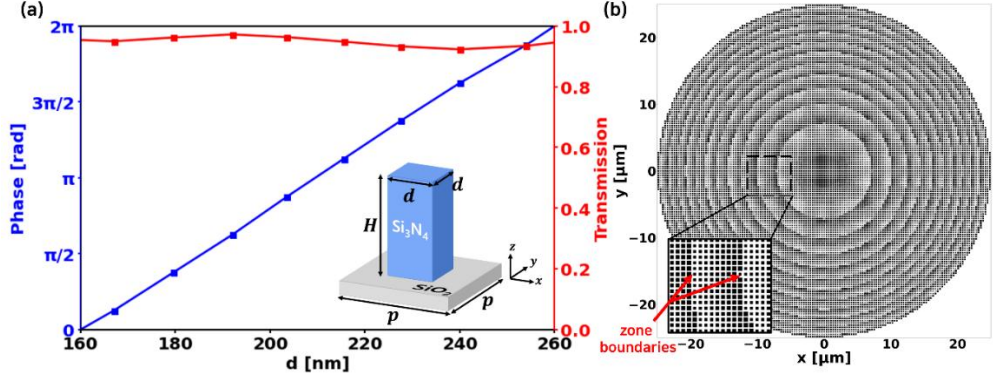

Fig. 1. (a) Phase delay and transmission of the $Si_3N_4$ meta-atom versus pillar diameter $d$, showing nearly full $2\pi$ phase coverage while maintaining high transmission. The inset illustrates the unit-cell geometry. (b) Top view of the corresponding metalens layout constructed by mapping the required phase profile onto the meta-atom library. The inset shows a magnified view of the meta-atom arrangement, where representative Fresnel-zone boundaries are indicated by arrows.

First, we compared the focal behaviors of a refractive lens and a metalens under variations in material refractive index. For a fair comparison, both devices were designed with the same diameter $D = 50$ μm and similar target focal length $f_{\text{target}} = 40$ μm, and were evaluated under identical FDTD simulation settings at a single wavelength $\lambda_0 = 633$ nm.

The metalens was implemented using a $Si_3N_4$ pillar meta-atom platform on a $SiO_2$ substrate. The meta-atoms were arranged on a square lattice with period $p = 340$ nm, and each pillar had a fixed height $H = 1500$ nm. The required focusing phase was defined by the hyperbolic phase profile

$$\phi_{\text{target}}(r) = -\frac{2\pi}{\lambda_0}\left(\sqrt{r^2 + f_{\text{target}}^2} - f_{\text{target}}\right), \tag{1}$$

which describes the ideal phase profile required to focus an incident plane wave at f_target [16,28]. To realize this target phase distribution, we constructed a meta-atom library by sweeping the pillar diameter d and calculating the corresponding phase delay and transmission at the nominal refractive index $n_0 = 2.0394$, as shown in Fig. 1(a). The phase increases nearly monotonically with $d$ while maintaining high transmission over the selected design range, enabling full $2\pi$ phase coverage. Based on this library, each lattice site was assigned the pillar diameter whose phase response best matched the required local phase, thereby generating the metalens layout shown in Fig. 1(b).

Meanwhile, a refractive convex-plano lens was designed under the same aperture and focal-length conditions as the metalens. The lens consisted of a spherical incident surface and a planar exit surface, with the curvature set by the thin-lens geometrical relation $R = f_{\text{target}}(n_0 - 1)$ at the nominal refractive index $n_0 = 2.0394$. This geometry was then kept fixed while only the refractive index was varied, allowing the effect of refractive-index variation on the focal behavior to be directly compared with that of the metalens under identical FDTD conditions.

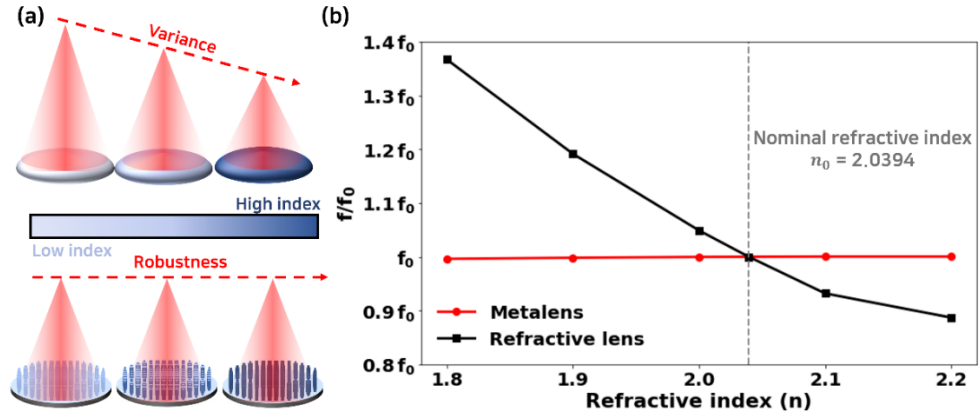

Fig. 2. Comparison of refractive-index-dependent focusing behaviors between a refractive lens and a metalens. (a) Schematic illustration of refractive index robustness in the metalens compared with the refractive lens. (b) Normalized best-focus position, $f/f_0$, versus refractive index $n$. The metalens shows an almost invariant best-focus position, whereas the refractive lens exhibits a pronounced monotonic shift. The dashed line marks the nominal refractive index, $n_0 = 2.0394$.

In both cases, the lens geometry and operating wavelength were fixed, while only the material refractive index was treated as an independent variable; the refractive index was swept from $n = 1.8$ to $n = 2.2$ around the nominal $Si_3N_4$ (silicon nitride) value $n_0 = 2.0394$ [30,31] to assess the impact of refractive index variations. Here, $f$ denotes the best-focus axial position, defined as the axial location that maximizes the on-axis intensity. This trend is also illustrated schematically in Fig. 2(a) and plotted quantitatively in Fig. 2(b). As shown in Fig. 2(b), the refractive convex-plano lens exhibits a monotonic shift of the best-focus position toward shorter distances as the refractive index increases, consistent with its enhanced refractive power [32,33]. In contrast, the meta-atom-based metalens shows a markedly different response: the best-focus position remains nearly unchanged despite variations in the material refractive index and is largely preserved even under a large refractive index sweep ($\Delta n/n_0 \approx \pm 10\%$). These results provide clear evidence that the metalens exhibits a fundamentally different and highly robust focusing behavior against refractive-index variation.

### 3.2 Physical origin of refractive index robustness

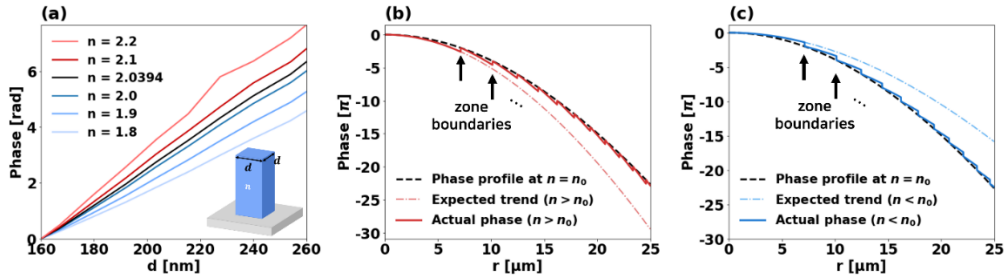

Fig. 3. Refractive-index-dependent modification of the meta-atom phase response and the resulting metalens phase profile. (a) Transmission phase as a function of pillar diameter $d$ for several refractive indices around the nominal value $n_0 = 2.0394$, showing that the phase response varies systematically with refractive index. (b) For $n > n_0$, comparison of the phase profile at $n = n_0$, the expected trend under refractive-index variation, and the actual phase realized by the original metalens design. (c) For $n < n_0$, comparison of the phase profile at $n = n_0$, the expected trend under refractive-index variation, and the actual phase realized by the original metalens design. In (b) and (c), the abrupt phase resetting in the solid curves originates from the zone boundaries defined in the original metalens design at $n = n_0$.

To clarify the physical mechanism of the observed robustness, we next examine how refractive index variations modify the phase response of the constituent meta-atoms and, consequently, the radial phase profile of the metalens. Fig. 3(a) shows the simulated transmission phase versus pillar diameter $d$ for several refractive indices around the nominal value $n_0 = 2.0394$. As the refractive index increases, the phase variation with pillar diameter becomes steeper, whereas it becomes flatter when the refractive index decreases. This trend is also qualitatively consistent with the approximate phase accumulation relation:

$$\phi \approx \frac{2\pi}{\lambda_0} n_{\text{eff}} H \qquad (2)$$

for a fixed pillar height $H$, a larger refractive index generally leads to a larger effective optical path length and hence a stronger phase accumulation, while a smaller refractive index leads to a weaker one. This means that even for an identical pillar geometry, the phase delay imparted by each meta-atom is directly altered by the refractive index. Therefore, when a metalens designed at $n_0$ is subjected to a different refractive index, its implemented phase profile cannot remain exactly the same.

This perturbation can be understood more clearly from Figs. 3(b, c), which compare the phase profile used in the original design at $n = n_0$, the expected phase trend under refractive index variation, and the actual metalens phase. In each panel, the dashed black curve denotes the phase profile used in the original design at $n_0$. The blue and red dash-dotted curves schematically represent the phase trends expected from the index-induced change in local phase accumulation: it becomes steeper for $n > n_0$ and flatter for $n < n_0$. In the actual metalens, however, the phase does not simply become a steeper or flatter version of this continuous trend. Instead, the solid line in Fig. 3(b) and (c), which denotes the actual metalens phase, undergoes a $2\pi$ jump [29] at each nominal zone boundary established during the original design at $n_0$. Because of these zone-boundary jumps, the actual metalens phase remains much closer to the nominal phase profile than to the continuous expected phase trend, while the phase slope within each zone is modified by the refractive-index variation. As a result, the actual metalens phase is expressed as piecewise segments with nearly preserved jump locations, rather than following the dash-dotted curve itself. Consequently, because the actual phase profile remains close to that of the nominal design, the metalens exhibits strong robustness against refractive index variations, and the best-focus position remains nearly unchanged.

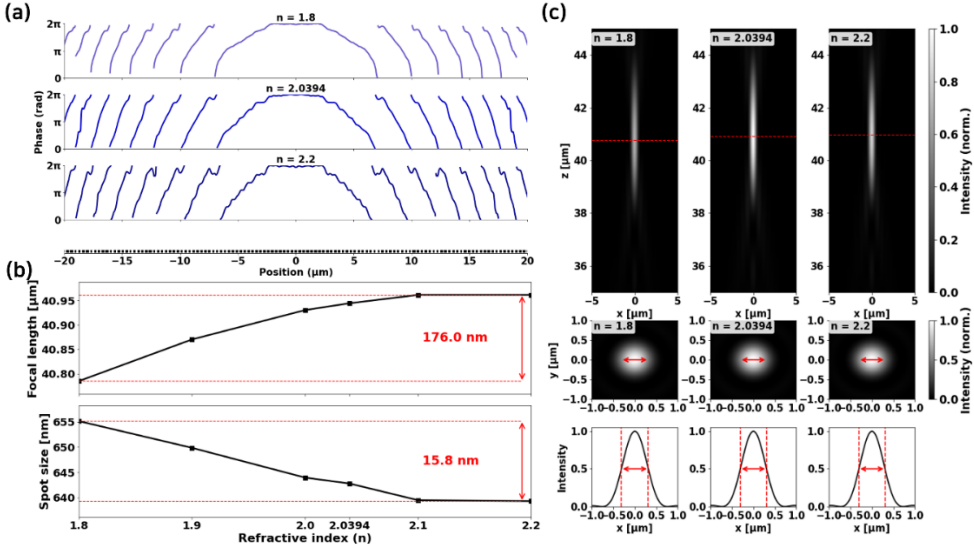

Fig. 4. Wrapped phase distributions and focusing performance of the metalens under refractive-index variations. (a) Wrapped radial phase profile at $n = 1.8, 2.0394,$ and $2.2$. (b) best-focus position and spot size versus refractive index. (c) $xz$ intensity maps and focal-plane intensity profiles at each refractive index.

To verify that the observed robustness indeed originates from the $2\pi$ phase jumps at the zone boundaries, we analyzed the phase distributions across the metalens for three representative refractive-index cases by the full FDTD simulation. Specifically, the near-field phase distributions were extracted from a monitor located at $z = 1.02H$ (30 nm above the metalens). Using $n_0 = 2.0394$ as the reference, we compared the wrapped-phase distributions at the extreme indices $n = 1.8$ and $n = 2.2$, and found that the overall phase distributions remain nearly identical despite the index variation, as shown in Fig. 4(a). This preserved phase distribution contrasts with the refractive lens case (without zone boundaries), where the phase distribution changes much more noticeably under refractive-index variations (see Supplement 1). Fig. 4(b) quantitatively presents the metalens' best-focus shift and spot size variation over the refractive index sweep ($\Delta n/n_0 \approx 10\%$). Calculated by a near-to-far-field projection (see Methods), the best-focus shift is confined within 176 nm ($\Delta f / f_0 \approx 0.4\%, \sim 0.28\lambda_0$), while the spot size variation remains within 15.8 nm ($\Delta w / w_0 \approx 2.5\%, \sim 0.025\lambda_0$). Notably, Fig. 4(c) visually corroborates these trends, showing that the $xz$ intensity maps and focal-plane intensity profiles remain nearly unchanged across the refractive index sweep. Indeed, the peak intensity decreases by approximately 30% at $n = 1.8$, but remains stable with less than 2% variation over the range $n = 2.0$ to 2.1. Despite these variations in light intensity, the geometric focusing characteristics—best-focus position and spot size—remain remarkably invariant throughout the entire refractive index sweep. These results show that the metalens maintains its focusing performance under refractive index variations, indicating focusing robustness. Importantly, this trend is not limited to a single operating condition and is consistently observed across different NA regimes (e.g., low-NA and high-NA cases) (see Supplement 1).

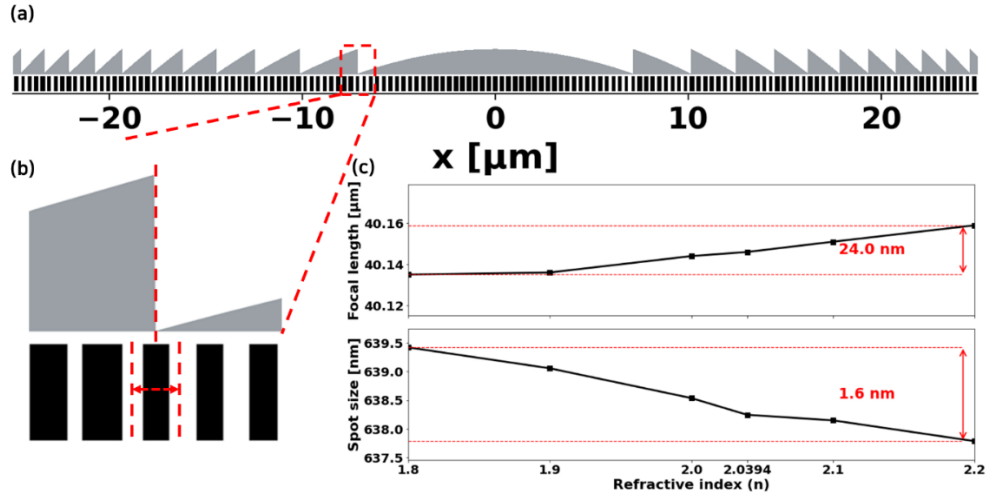

Fig. 5. (a) Geometric comparison between the Fresnel lens and the metalens. (b) Magnified view illustrating the alignment between the Fresnel zone boundary and the corresponding metalens zone boundary. (c) Best-focus position and spot size of the Fresnel lens as functions of refractive index.

To support our hypothesis that zone boundaries are a key structural origin of robustness, we transformed a conventional refractive lens into a Fresnel lens with physically well-defined zone boundaries. If zone boundaries indeed underlie robustness, introducing zone boundaries into a continuous refractive lens should suppress the sensitivity of the lens to refractive-index variations. The Fresnel lens provides an ideal platform to test this hypothesis. The conversion of a refractive lens into a Fresnel lens effectively folds the continuous lens surface into concentric steps by removing thickness corresponding to an integer multiple of $2\pi$ phase delay. Consequently, the resulting physical zone boundaries correspond to the phase-reset boundaries of the metalens profile (Figs. 5(a, b)).

While classical diffraction theory analytically predicts that the focal length of a Fresnel lens is governed by zone radii [32,33] rather than by the material refractive index of the diffractive element [34], relying solely on theoretical approximations is insufficient for a direct quantitative comparison with full wave metalens results. Therefore, to ensure a fair comparison, we conducted FDTD simulations for the Fresnel lens under numerical conditions identical to those used for the metalens, thereby explicitly confirming its optical behavior through full wave analysis.

As shown in Fig. 5(c), the Fresnel lens exhibits near-ideal robustness even when the refractive index is swept over a wide range of $\Delta n = 0.4$. Over this sweep, the best-focus position shift remains as small as 24.0 nm and the spot size variation is limited to 1.6 nm. In other words, although the original refractive lens shows substantial focal variation under refractive-index changes, such variation is almost completely suppressed in the Fresnel lens once explicit zone boundaries are introduced. This is also consistent with the classical diffraction-based expectation that the focusing behavior of a Fresnel lens is governed primarily by its zone boundaries, and further supports our hypothesis. By comparison, although the overall focusing behavior of the metalens remains stable, it shows slightly larger deviations under the same refractive index variation. This is because the zone boundaries are physically well-defined in the Fresnel lens, whereas those of the metalens are less distinct because the phase profile is implemented in a discretized manner.

## 3.3 Quasi-scale-invariant focusing behavior

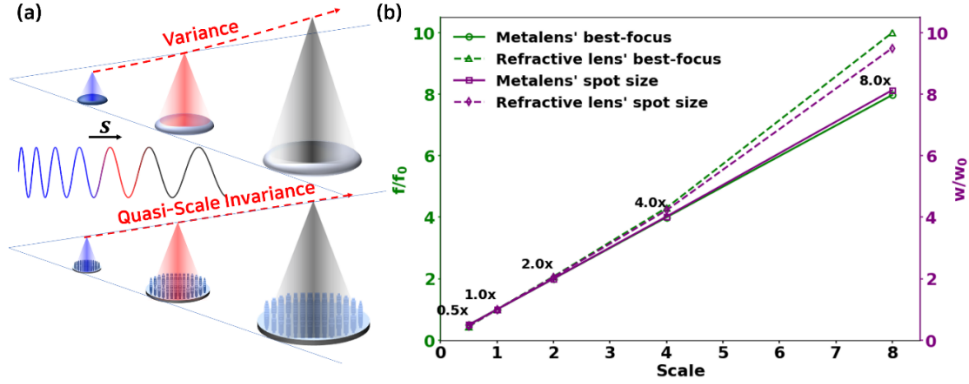

Fig. 6. Quasi-scale-invariant focusing behavior of the metalens. (a) Conceptual illustration comparing scale-dependent variance in the refractive lens (top) and quasi-scale-invariant focusing in the metalens (bottom) under uniform geometric scaling by a factor $s$. (b) Normalized best-focus position $f/f_0$ and spot size $w/w_0$ versus the scale factor for the metalens (solid lines) and refractive lens (dashed lines). The metalens preserves nearly linear scaling, whereas the refractive lens exhibits noticeable deviation at larger scale factors.

Moreover, this robustness, which preserves focusing under refractive index variations, can also extend to uniform geometric scaling of the entire geometric structure, suggesting a predictable scaling trend in the focusing response. This scaling perspective is closely connected to the well-known scale invariance in electromagnetism [35,36]. In an ideal non-dispersive medium where relative permittivity and permeability are frequency-independent, uniformly scaling all geometric lengths by a factor $s$ together with the wavelength leads one to expect the same optical functionality reproduced in a scaled form. Under such ideal conditions, geometric scaling yields deterministic scaling relations for the optical response. However, real optical materials are dispersive, so the refractive index depends on wavelength, and the exact scale invariance expected under ideal conditions is therefore generally broken [37]. Even if the geometry and the wavelength are scaled together by a factor $s$, the material refractive index does not remain identical, so the scaled system is not strictly equivalent to the original one. Nevertheless, because the metalens exhibits robustness in its focusing behavior against refractive index variations, the focusing functionality can remain a good approximation under scaling even when dispersion prevents the ideal scaling condition from being met exactly. In other words, exact scale invariance is broken in a conventional refractive lens made of dispersive materials, and its behavior therefore varies with scale. By contrast, the metalens can retain a systematic scaling trend in its focusing functionality, aided by its robustness against refractive index variations, leading to quasi-scale-invariant behavior. This contrast is summarized in the conceptual illustration in Fig. 6(a).

Table 1. Scaled wavelength $\lambda = s\lambda_0$ and the corresponding dispersive refractive index $n(\lambda)$ of the $Si_3N_4$ pillar material used in the quasi-scale-invariance test.

|  | $s = 0.5$ | $s = 1.0$ | $s = 2.0$ | $s = 4.0$ | $s = 8.0$ |
| --- | --- | --- | --- | --- | --- |
| Wavelength $\lambda$ [nm] | 316.5 | 633 | 1266 | 2532 | 5064 |
| Refractive index $n$ of $Si_3N_4$ pillar | 2.1678 | 2.0394 | 2.0046 | 1.9661 | 1.8315 |

To quantify the quasi-scale-invariant focusing behavior, the nominal designs of both the metalens and the refractive lens were uniformly scaled over $s = 0.5$ to 8. The scaling was applied self-consistently not only to each lens geometry but also to the wavelength $\lambda = s\lambda_0$, the FDTD simulation region, and the size/placement of the field monitors, enabling a consistent comparison across the scaled systems. At each scaled wavelength, we also accounted for the dispersive refractive index $n(\lambda)$ of the $Si_3N_4$ pillar material, summarized in Table 1 (the full scaling parameters are provided in Table S3 of Supplement 1). As shown in Fig. 6(b), the normalized best-focus position $f/f_0$ and spot size $w/w_0$ of the metalens (solid lines) vary nearly linearly with $s$, indicating a predictable scaling trend in the focusing behavior. By contrast, the refractive lens (dashed lines) exhibits noticeable deviation at larger scale factors. Quantitatively, over the full scaling range from $s = 0.5$ to 8, the maximum deviations from ideal linear scaling are only 0.3% for $f/f_0$ and 1.7% for $w/w_0$ in the metalens. In the refractive lens, the corresponding deviations increase to 25.0% and 18.7%, respectively.

Beyond revealing a scaling trend, this quasi-scale-invariant behavior of the metalens suggests practical implications. For example, this approach may be particularly useful for ultraviolet metalenses. In ultraviolet refractive systems, absorption-induced heating can cause time-dependent focal drift [38], whereas recent studies suggest that properly engineered metalenses can exhibit reduced thermal drift under intense irradiation [9,39]. However, practical implementation in the ultraviolet remains challenging because transparent material platforms are limited and the required nanostructures become extremely small. In such cases, a longer-wavelength scaled counterpart could serve as a practical surrogate for preliminary validation before implementation at shorter wavelengths. It also indicates the potential to share the same metalens design across different wavelength bands. In this sense, a design validated at one wavelength may be transferred to other spectral regions through scaling, reducing the need for repeated re-design while preserving a predictable focusing functionality.

## 4. Conclusion

In conclusion, this work goes beyond simply reporting the focusing robustness of a metalens under refractive-index variations by identifying its physical origin. Although refractive-index fluctuations modify the phase response of individual meta-atoms, the overall phase profile remains close to that of the nominal design because the zone boundaries defined by the original design continue to play the dominant structural role. As a result, the focusing behavior is only weakly perturbed, and the best-focus position and spot size remain largely stable over a broad index range. In this sense, the observed robustness is not accidental, but is rooted in the way the designed phase profile is structurally maintained even under local phase-response changes. We further show that this robustness is also connected to a quasi-scale-invariant focusing trend under uniform geometric scaling in realistic dispersive conditions, where strict scale invariance is not expected. Overall, these results show that refractive-index robustness in metalenses can be understood through the role of zone boundaries in preserving the designed phase profile, and provide broader physical insight into what governs robustness in metasurface optics.

**Supplementary information**

**1. Refractive Index-Dependent Phase Profiles of a Convex-Plano Refractive Lens**

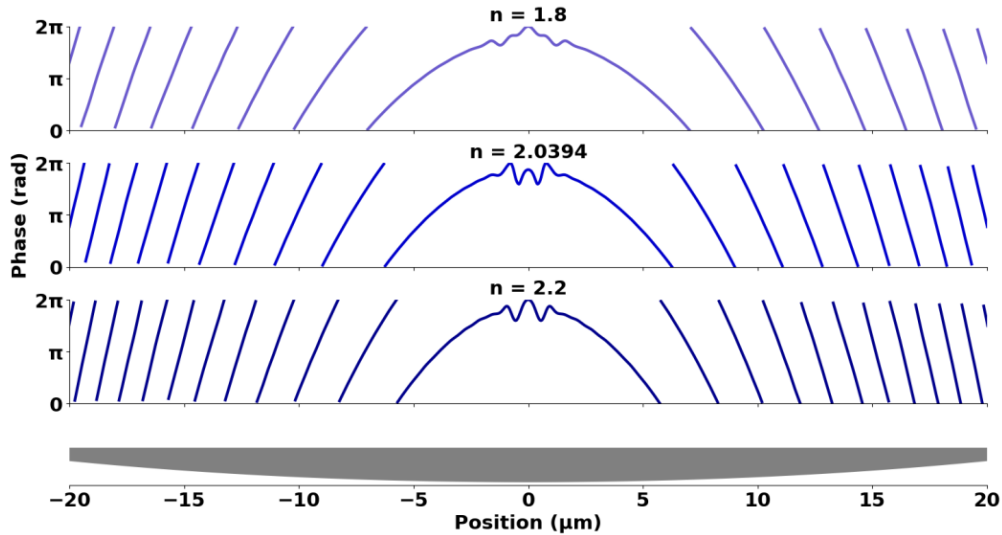

Fig. S1. Wrapped phase profiles of a convex–plano refractive lens for refractive indices $n = 1.8$, 2.0394, and 2.2. As the refractive index varies, the phase gradient changes, while the overall phase curvature remains similar.

To provide a comparative reference, we analyzed a conventional convex–plano refractive lens under refractive index variation. For a fair comparison with the metalens, the lens geometry was fixed, and the wrapped phase profiles were extracted from full wave simulations for $n = 1.8$, 2.0394, and 2.2. As shown in Supplementary Fig. S1, the phase gradient varies significantly with refractive index. In particular, the phase profile becomes steeper as the refractive index increases and flatter as it decreases. This trend indicates that, in a conventional refractive lens (without zone boundaries), the accumulated phase is directly modified by refractive index variation. Consequently, unlike in the metalens, the phase profile is not preserved, leading to changes in the effective wavefront and corresponding shifts in the best-focus position (see Fig. 2(b)).

## 2. Additional NA cases

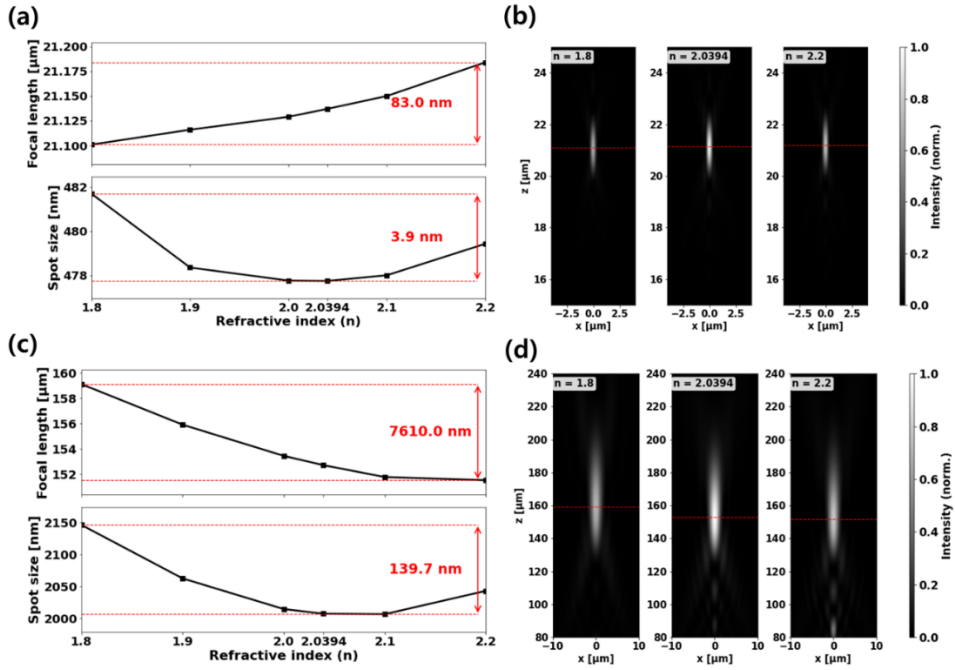

Fig. S2. Additional NA cases. (a,b) High-NA metalens ($D = 50$ μm, $f \approx 20$ μm, $NA \approx 0.78$) and (c,d) Low-NA metalens ($D = 50$ μm, $f \approx 160$ μm, $NA \approx 0.15$). (a,c) best-focus position and spot size (FWHM) versus refractive index $n$. (b,d) normalized x-z intensity map at $n = 1.8, 2.0394$, and 2.2 (red dashed lines: best-focus positions).

To assess the generality of the refractive index robustness discussed in the main text, we performed the same refractive index sweep for two additional metalens designs with different NA. The high-NA metalens has $D = 50$ μm, $f \approx 20$ μm, and $NA \approx 0.78$ (Fig. S2 (a,b)), while the low-NA metalens has $D = 50$ μm, $f \approx 160$ μm, and $NA \approx 0.15$ (Fig. S2 (c,d)). For the high-NA case, both the best-focus shift and the spot size change remain small over $n = 1.8$ to 2.2 ($\Delta f = 83.0$ nm and $\Delta w = 3.9$ nm), and the normalized x-z intensity maps are nearly unchanged, indicating preserved focusing behavior. For the low-NA case, the absolute variations become large ($\Delta f = 7.61$ μm and $\Delta w = 139.7$ nm), however, the normalized x-z intensity map in Fig S2(d) shows that the overall focal intensity morphology remains similar across the index sweep. These results suggest that the robustness is not confined to a specific NA regime and persists across both high- and low- NA designs.

## 3. Scaling protocol for quasi-scale-invariant focusing

This section summarizes the definition of uniform geometric scaling used in the main text and clarifies the scope of parameters that were jointly scaled in the FDTD simulations. We define the scaling factor $s$ relative to the baseline design at $s = 1$. The wavelength and all length-type geometric parameters were scaled as $\lambda(s) = s\lambda_0$ and $L(s) = sL_0$. Since the speed of light $c$ is fixed, scaling only spatial lengths by $s$ changes the corresponding propagation time scale. To maintain space–time consistency, time-related simulation parameters were also scaled as $t(s) = st_0$. Material dispersion was explicitly included. At each $\lambda(s)$, the refractive indices of the $Si_3N_4$ pillar and $SiO_2$ substrate were updated to the corresponding wavelength-dependent values, enabling an assessment of scaling trends under realistic dispersive conditions where strict scale invariance is not assumed. The $Si_3N_4$ refractive index values used at each wavelength are summarized in Table 1 of the main text.

The list below summarizes the parameters that were jointly scaled under $L(s) = sL_0$ and $t(s) = st_0$.

1. Geometry (design): For the metalens case, period of unit cell, height of $Si_3N_4$ pillars, lens diameter $D$, diameter of meta-atoms $d$, thickness of substrate; for the refractive lens case, lens diameter $D$, center thickness, and radius of curvature.

2. Simulation conditions: FDTD region span $L_x, L_y, L_z$, size and position of field monitor for near-field data, size and position of source (source-to-metalens distance), z-scan range for best-focus search, z-scan step size, focal-plane monitor size, and mesh cell size (set relative to $\lambda(s)$)

3. Total simulation time.

The table below lists representative parameters jointly adjusted under the scaling applied in this work.

Table S3. Summary of representative scaled parameters.

| Scale $s$ | Wavelength $\lambda$ [nm] | Lens diameter [μm] | Refractive index of $SiO_2$ substrate | Height of $Si_3N_4$ pillar [μm] | Period of unit cell [μm] |
|---|---|---|---|---|---|
| 0.5 | 316.5 | 25 | 1.4835 | 0.75 | 0.17 |
| 1 | 633 | 50 | 1.457 | 1.5 | 0.34 |
| 2 | 1266 | 100 | 1.4473 | 3 | 0.68 |
| 4 | 2532 | 200 | 1.4292 | 6 | 1.36 |
| 8 | 5064 | 400 | 1.3364 | 12 | 2.72 |